\documentclass{aa} 
\usepackage{graphicx}
\usepackage{txfonts}

\def\sn{\hbox{S/N}}  
  
\def\vsin{\hbox{$v \sin i$}}

\def\kms{\hbox{km\,s$^{-1}$}}

\def\degr{\hbox{$^\circ$}}

\def\kis{\hbox{$\chi^2$}}   
\def\kisr{\hbox{$\chi^2_{\rm r}$}}

\def\msun{\hbox{$M_\odot$}}   
\def\rsun{\hbox{$R_\odot$}}

\begin{document} 

\title{A decade-long magnetic monitoring of Vega}
\titlerunning{Decade-long magnetic monitoring of Vega}

\author{
P. Petit\inst{1}
\and
T. B\"ohm\inst{1}
\and
C.P. Folsom\inst{2}
\and
F. Ligni\`eres\inst{1}
\and
T. Cang\inst{1,3}
}
         
\institute{
Institut de Recherche en Astrophysique et Plan\'etologie, Universit\'e de Toulouse, CNRS, CNES, 14 avenue Edouard Belin, 31400 Toulouse, France 
\and 
Tartu Observatory, University of Tartu, Observatooriumi 1, Toravere, 61602 Tartumaa, Estonia
\and
Department of Astronomy, Beijing Normal University, Beijing 100875, P. R. China}
             
\date{}

\abstract
{}
{The very weak magnetic field detected at the surface of Vega hints at a widespread population of weakly magnetic stars of A and B spectral types. We contribute here to gather more clues about the origin of this magnetism by investigating the long-term stability of the field geometry of this prototypical star.}
{We use spectropolarimetric data collected as part of a long-term campaign, with more than 2,000 observations spread between 2008 and 2018. Using various sub-sets extracted from the whole time series, we reconstruct several maps of the large-scale surface magnetic field.}
{We confirm that the polarimetric signal is modulated according to a $\sim 0.68$~d period, which we interpret as the stellar rotation period. The surface magnetic field is organized in a complex geometry. We confirm the existence of a very localized, polar magnetic spot previously reported for Vega, with a radial field strength of about -5~G. We show that the surface of the star is also covered by a dipole, with a polar strength close to 9~G and a dipole obliquity close to 90\degr{}. Both magnetic structures are remarkably stable over one decade. The available data suggest that smaller-scale magnetic spots may not be limited to the polar region, although the poor reliability of their reconstruction does not allow us to firmly conclude about their temporal evolution.}
{}


\maketitle

\section{Introduction}

The very first magnetic field reported on a star other than the Sun was detected on a main sequence A-type star more than seven decades ago \citep{1947ApJ...105..105B}. Following this major discovery, it was soon understood that main sequence stars of intermediate mass with chemical peculiarities (Ap and Bp stars) host strong magnetic fields, and it is now established that this population amounts to about 3\% of stars with masses around 1.5~\msun{}, and up to 10\% of stars with masses between 3 et 3.8 \msun{}\ \citep{2019MNRAS.483.3127S}. In most cases, their surface magnetic field is dominated by simple dipoles, although high precision spectropolarimetric observations highlight more complex surface distributions of magnetic fields (e.g. \citealt{kochukhov04}). These magnetic configurations seem to be very stable, with no intrinsic variability reported at the spatial scales investigated so far \citep{silvester14}. 

The distribution of dipolar strengths among Ap/Bp stars features a lower limit at around 300~G \citep{auriere07}, and observing campaigns aimed at detecting magnetic fields outside of this specific population remained inconclusive for decades (e.g. \citealt{shorlin02,2006A&A...451..293W,auriere10,makaganiuk11} for a few recent attempts). This frustrating situation changed when an ultra-deep spectropolarimetric campaign resulted in the detection of a weak magnetic field on Vega \citep{lignieres09}. The measured surface field was much weaker than any previous detection in intermediate-mass stars, with a maximal magnetic strength reaching seven gauss locally \citep{petit10,2014A&A...568C...2P}. This first successful attempt was later repeated with similar detections obtained for several bright Am stars \citep{petit11,blazere16a}, suggesting that weak magnetic fields may be a common feature of A and B stars without Ap/Bp characteristics. This speculation, based on a small number of objects, is supported by the observation, in space-borne photometric surveys, of non-pulsational variability in intermediate-mass stars without strong magnetic fields \citep{2017MNRAS.467.1830B,2019MNRAS.490.2112B}.

A couple of stars in the same mass regime were also reported to host surface fields with a strength intermediate between Vega-like objects and chemically peculiar stars \citep{blazere16b,2020MNRAS.492.5794B,alecian16}. Such stars have to be relatively rare. Otherwise, their siblings would have been spotted before. As of today,  it remains difficult to decide whether they belong to the lower tail of the strong field distribution, or to the upper tail of the weak-field population, or if they are the product of a more atypical evolutionary channel.

This preliminary exploration of a new class of magnetic A-type stars prompted the development of scenarios and models aimed at understanding the origin and evolution of these weak magnetic fields. Some of the proposed scenarios are based on the assumption that the observed fields are the product of the slow decay of stronger fields by instabilities  \citep{auriere07,2013MNRAS.428.2789B,2015A&A...580A.103G,2020ApJ...900..113J}, or the product of an internal dynamo \citep{cantiello2011,2019ApJ...883..106C}.  

Little is known, so far, about the stability of such weak surface fields, with the exception of hints for short-term magnetic variability reported for the Am star Alhena \citep{2020MNRAS.492.5794B}. As the prototype of weakly magnetic A-type stars, Vega is the focus of the present study, dedicated to the long-term monitoring of its photospheric field. The spectral type of Vega is A0 \citep{2003AJ....126.2048G}, and its photospheric chemical composition is affected by a depletion of iron peak elements, usually referred to as a $\lambda$~Bo\"otis chemical anomaly \citep{1990ApJ...363..234V}. This chemical property may come from the recent accretion of material from its debris disc observed in the near-infrared \citep{2005ApJ...628..487S}. Apart from this weakly expressed specificity, Vega can be considered a good representative of early A stars with normal surface abundances, making it an interesting object to study stellar magnetism in the absence of chemical peculiarities. The spin axis of Vega is seen at a low inclination angle of around $7$\degr{}, as shown by interferometry \citep{2006ApJ...645..664A} and spectroscopy  \citep{2008ApJ...678..446T,2021MNRAS.505.1905T}. Its relatively small projected rotational velocity \vsin{}~$\sim 22$~\kms\ (compared to other normal A-type star, \citealt{2012A&A...537A.120Z}) is, therefore, hiding a high equatorial velocity of $195 \pm 15$~\kms\ \citep{2021MNRAS.505.1905T}.

In the following sections, we first present our observational material, detail the methods used to extract Zeeman signatures, obtain an estimate of the stellar rotation period using the polarized signal, reconstruct the large-scale magnetic field geometry of the star, and discuss our results in the light of other published works. 

\section{Spectropolarimetric time series}

\subsection{Instruments}

Our observations were carried out with the NARVAL and ESPaDOnS twin instruments \citep{auriere03,2006ASPC..358..362D}. The first one is installed at the Télescope Bernard Lyot (Observatoire du Pic du Midi, France) and therefore benefits from a 2~m reflector. The second one is operated at the Canada-France-Hawaii Telescope (Mauna Kea Observatory, Hawaii), with the advantage of a 3.6~m primary mirror. Apart from this difference in collecting area, both instruments are based on the same design. They are high resolution \'echelle spectropolarimeters ($R \sim 65,000$ in polarimetric mode), covering the whole optical domain in a single exposure (370 to 1,000 nm, with only small spectral gaps between the red-most spectral orders). They can record any Stokes parameter, with polarized Stokes parameters being built from four sub-exposures recorded at different rotation angles of the Fresnel rhombs. In practice, our data are all obtained in the Stokes V configuration, as the amplitude of Zeeman signatures is the largest in circular polarization \citep{1992soti.book...71L,kochukhov04}. All data were reduced using the automated LibreEsprit package \citep{donati97}. The final product of the reduction is a normalized Stokes I (intensity) spectrum, and the corresponding Stokes V spectrum. Using a different combination of the sub-exposures used to build the Stokes V spectrum, the data reduction pipeline provides us with an additional spectrum that should contain pure noise. We use this so-called Null spectrum as a sanity check that spurious polarized signatures of instrumental or stellar origin do not contaminate our measurements.

\subsection{Spectrolarimetric observations}

\begin{table*}
\caption{Journal of observations. For data obtained in 2009, we first provide information about the NARVAL and ESPaDOnS runs taken separately, then about the two subsets merged together. The last line is for the combination of all available data.} 
\centering
\begin{tabular}{c c c c c c c c}
\hline
Year & Instrument & Nobs & HJD min. & HJD max. & HJD range & $\langle S/N \rangle$ & $\sigma(S/N)$ \\
\hline
2008 & NARVAL & 259 & 2454673.47 & 2454676.63 & 3.16 & 22041 & 3353\\
2009 & NARVAL & 238 & 2455005.66 & 2455139.24 & 133.58 & 21356 & 6423\\
2009 & ESPaDOnS & 321 & 2455083.74 & 2455085.95 & 2.21 & 20822 & 3293\\
2009 & NAR+ESP & 559 & 2455005.66 & 2455139.24 & 133.58 & 21050 & 4885\\
2010 & NARVAL & 615 & 2455393.37 & 2455425.60 & 32.23 & 28180 & 2405\\
2014 & NARVAL & 301 & 2456830.54 & 2456865.51 & 34.96 & 20773 & 5049\\
2018 & NARVAL & 568 & 2458331.33 & 2458337.63 & 6.29 & 21410 & 3687\\
\hline
2008-2018 & NAR+ESP & 2302 & 2454673.47 & 2458337.65 & 3664.16 & 23118 & 4974\\ 
\hline
\end{tabular}
\label{tab:journal}
\end{table*}

The data were collected during six different observing campaigns, spread from 2008 to 2018. NARVAL data have been collected for five epochs, while ESPaDOnS observations were all gathered during a single run (Tab. \ref{tab:journal}). Due to the brightness of the target, exposure times were kept very short (from 4 to 12 sec for each of the four sub-exposures constituting a Stokes V sequence). A small fraction of the observations occasionally leads to poor \sn, generally because of transitory failures of the auto-guiding. All observations with abnormally small \sn\ (\sn\ of LSD profiles below 10,000, see Sect. \ref{sec:lsd}) were removed from our list, as well as a few spectra with abnormal radial velocities (more than 1 \kms\ away from the expected radial velocity). The data set retained after this basic filtering is constituted of 2302 spectra. All reduced data can be accessed through the PolarBase archive \citep{2014PASP..126..469P}.

\section{Zeeman signatures}
\label{sec:lsd}

\begin{figure} 
\includegraphics[width=9.5cm]{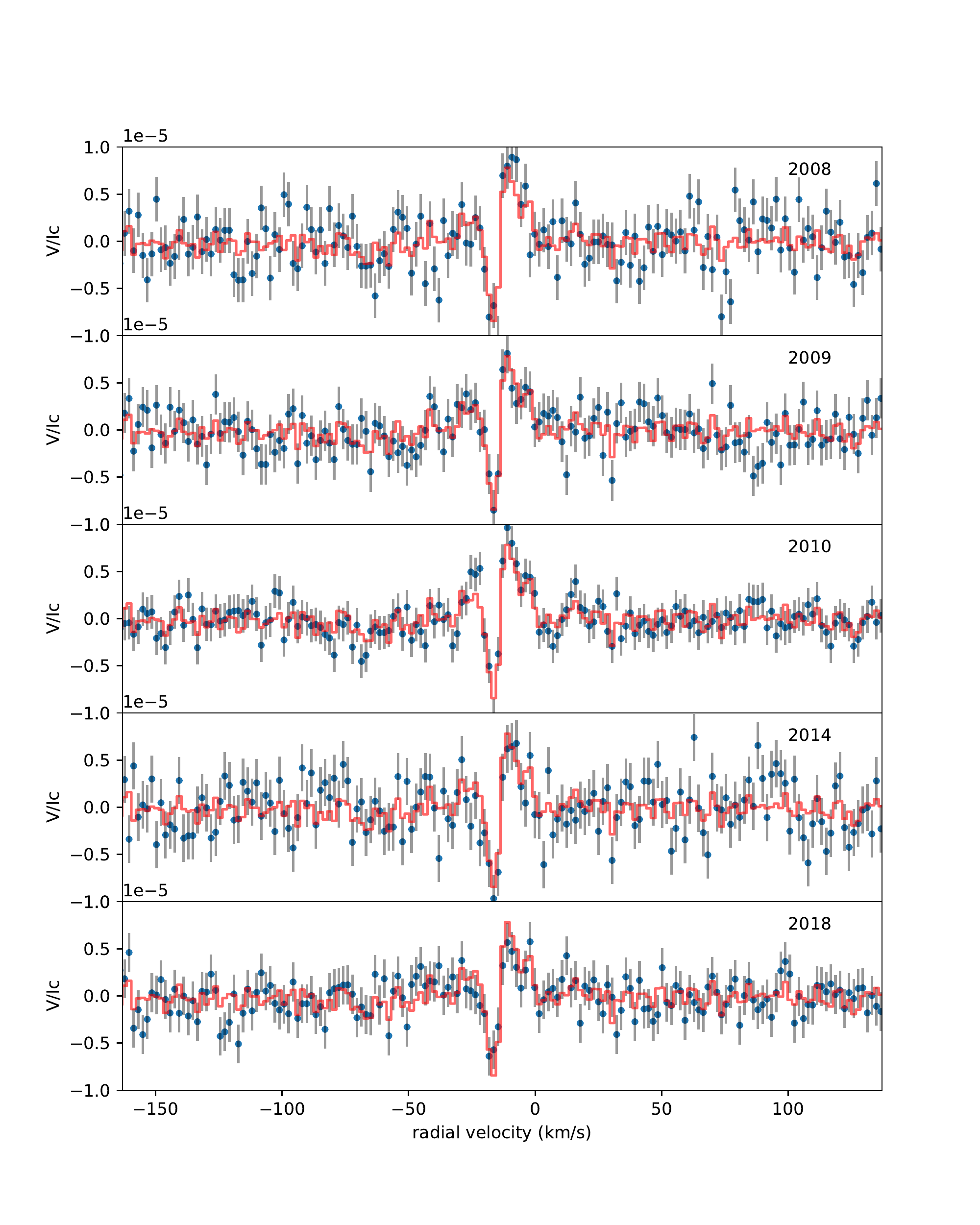} 
\caption{Averaged Stokes V LSD pseudo-profiles. Observations for each observing epoch are plotted as blue symbols, with the year of observation indicated in the right upper part of the plots. For each epoch, we over-plot in red the signature obtained by averaging all available data together.}
\label{fig:stokes}
\end{figure}

The reduced polarized spectra do not display any detectable polarized signatures in individual spectral lines. As widely done in similar situations, we applied the Least-Square-Deconvolution method (LSD, \citealt{donati97,kochukhov10}) for a multiplex extraction of Zeeman signatures from a list of photospheric lines, ignoring portions of the spectra affected by telluric bands or very broad lines (e.g. Balmer lines). We use here the same line list as the one previously employed by \cite{lignieres09} and \cite{petit10}, resulting in a total of $\sim 1,100$ lines deeper than 1\% of the continuum in the LSD analysis. The outcome is a set of pseudo-line profiles with greatly enhanced \sn\ compared to the initial spectra, with an average \sn\ of 23,000 for LSD profiles in Stokes V. The normalization wavelength of the LSD profiles is taken equal to 650~nm, and their normalization Land\'e factor is equal to 0.95.

The lower noise level reached after this first step remained too large to get any detection of a polarized signature. As a consequence, we repeated the strategy proposed by \cite{lignieres09} and averaged together all LSD profiles of individual epochs, to further reduce the relative noise. The outcome of this simple procedure is illustrated in Fig. \ref{fig:stokes}. We repeatedly observe polarized signatures in our averaged Stokes V profiles, around the central radial velocity of the line profile. The radial velocity width of the signatures is smaller than \vsin{}, and they display a peak-to-peak amplitude of $\sim 2 \times 10^{-5}I_c$, where $I_c$ is the continuum intensity. The shape of the signature is the one expected with a field of negative polarity. We do not report any statistically significant variation of the signature shape over the course of the monitoring. The Null profiles, plotted in Fig. \ref{fig:null}, do not show any detection of spurious signatures.

By averaging together hundreds of observations, we combine a random distribution of rotation phases. If phase-dependent signatures are present in the data, they are likely averaged out in this procedure. We investigate these phase-dependent signatures in Sec. \ref{sec:maps}. The only remaining signatures are those produced by the axisymmetric component of the large-scale field. The detected signal, with its limited extent in radial velocity, is generated by a radial magnetic spot very close to the visible rotation pole. We see here that this polar spot, already discussed in \cite{petit10}, is probably able to survive over one decade, with no noticeable evolution of its field strength.    

\section{Magnetic mapping}

Stellar rotation is expected to modify the shape and amplitude of Zeeman signatures, following the succession of stellar longitudes facing the observer. This property is exploited by the Zeeman-Doppler Imaging method (ZDI hereafter, \citealt{semel89}), which offers a powerful approach to reconstructing the 2D distribution of magnetic fields on the stellar surface. We use here the Python ZDI code of \cite{2018MNRAS.474.4956F}, implementing a maximum-entropy inversion of a time-series of Stokes V LSD profiles. In this model, the magnetic field geometry is projected onto a spherical harmonics frame \citep{donati06}, with the axis of spherical harmonics being the same as the spin axis. We limit here the spherical harmonics expansion to modes with $\ell \le 10$, since the inversion is not improved by allowing for smaller spatial scales to be modeled. The stellar surface is divided into a grid of rectangular pixels, with each pixel producing a local absorption line modeled as a Voigt profile \citep{2018MNRAS.481.5286F}. This local intensity profile is translated into a Stokes V line profile through the weak field approximation. The parameters of the local Voigt function are tuned to match the shape of the observed LSD Stokes I profiles, imposing a standard deviation $\sigma = 1.2$~\kms\ and a Lorentzian width $\Gamma = 1.33$~\kms. We include in our model an oblate stellar shape, mostly following the approach detailed in \cite{2020A&A...643A..39C}. To compute the rotational flattening, the stellar mass is taken equal to 2.15 \msun{}, and the equatorial radius to $R_\mathrm{eq} = 2.8$ \rsun\ \citep{2010ApJ...708...71Y}. With a rotation period of $\sim 0.68$~d and following our simple Roche model of the surface shape, these parameters lead to a polar radius $R_\mathrm{p} \sim 2.03$ \rsun, slightly smaller than the one reported by Yoon. We also assume that the local surface temperature varies as $g_\mathrm{eff}^{b}$, with $g_\mathrm{eff}$ being the effective gravity. Contrary to \cite{2020A&A...643A..39C} where the approach of \cite{1967ZA.....65...89L} was adopted, here the gravity darkening coefficient $b$ was set according to the prescription of \cite{2016LNP...914..101R}. According to this model, $b$ is related to the equatorial and polar radii through the equation $b \approx 1/4 - 1/3\epsilon$, with $\epsilon = 1 - R_\mathrm{p}/R_\mathrm{eq}$. In our surface model, the brightness was assumed to vary as $g_\mathrm{eff}^{4b}$. We tuned the values of the mass, equatorial radius, and rotation within a reasonable range for each parameter taken separately and conclude that the consequences of gravity darkening on the magnetic maps remain modest (up to 10\% changes in field strength, and small latitudinal shifts of magnetic spots). In practice, all conclusions of this paper regarding the rotation period, field geometry, and field stability remain unchanged if a spherical geometry and no gravity darkening are implemented in the ZDI model. Apart from gravity darkening and limb darkening, the local surface brightness of the star was assumed to be uniform. This was considered to be a reasonable approximation in the case of Vega since line bumps due to its surface brightness spots are known to display a very small amplitude \citep{2015A&A...577A..64B,2017MNRAS.472L..30P}. The oblate stellar surface is  not taken into account in the spherical harmonics expansion of the magnetic field.

The mapping procedure closely follows the successive steps detailed by \cite{2020A&A...643A..39C}. We kept the ZDI input parameters of \cite{petit10}, with slightly adjusted values for the inclination angle and \vsin{}\ (equal to 6.4\degr{}\ and 21.6 \kms, respectively), following \cite{2021MNRAS.505.1905T}. The adopted rotation period is the best period extracted from the Stokes V profiles (0.688544~d, see the detailed discussion in Sec. \ref{sec:period}, as well as Tab. \ref{tab:periods}), and the phase origin is chosen at the Julian date 2456892.015.

Vega is a very challenging target for ZDI, due to polarimetric signatures smaller than the noise, making it very difficult to avoid some level of over-fitting of the data. In other words, some of the reconstructed features will likely be generated by the noise pattern. We take advantage of the independent data sets collected over the years to better separate actual magnetic features from the noise artifacts, by looking for surface features that show up consistently using various sub-sets. Although this situation makes the inversion procedure more tricky than in most ZDI studies, one advantage is that the outcome of our ZDI models is poorly sensitive to the exact values of input ZDI parameters, so that slightly different values of \vsin{}, of the inclination angle, or of the limb darkening coefficient, have a limited impact on the resulting magnetic map. This is not the case, however, for the rotation period that needs to be tightly constrained to accurately identify rotation phases over a timespan as long as a decade. The standard inversion procedure, involving maximum entropy regularization, consists in setting a target reduced \kis{}\ (\kisr\ hereafter) for the magnetic model and maximizing the map entropy at this specific \kisr{}. Arbitrary low \kisr{} levels are generally not achievable, and the smallest accessible values tend to over-fit the data. In our case, the stable Stokes V signatures shown in Fig. \ref{fig:stokes} suggest that the magnetic geometry may be mostly stable over the time span of our data set. Thanks to this prior information, we can take advantage of different independent subsets to run a series of ZDI models and look for consistent surface features. The content of the resulting map is quite sensitive to the target \kisr{}, depending on the different noise levels of different subsets. Facing this difficulty, we choose here to set a target entropy (i.e. a target average field strength, chosen here close to 5.5~G) instead of a target \kisr{} and let the code minimize the \kisr{}\ at this entropy value (following, e.g., \citealt{petit02}). By doing so, we assume that the average surface field strength is constant over a decade, which is supported by the stability of the averaged Stokes V profiles. The resulting \kisr\ values are given in Tab. \ref{tab:periods}.

\section{Rotation period search}
\label{sec:period}

\begin{figure} 
\centering
\includegraphics[width=9cm]{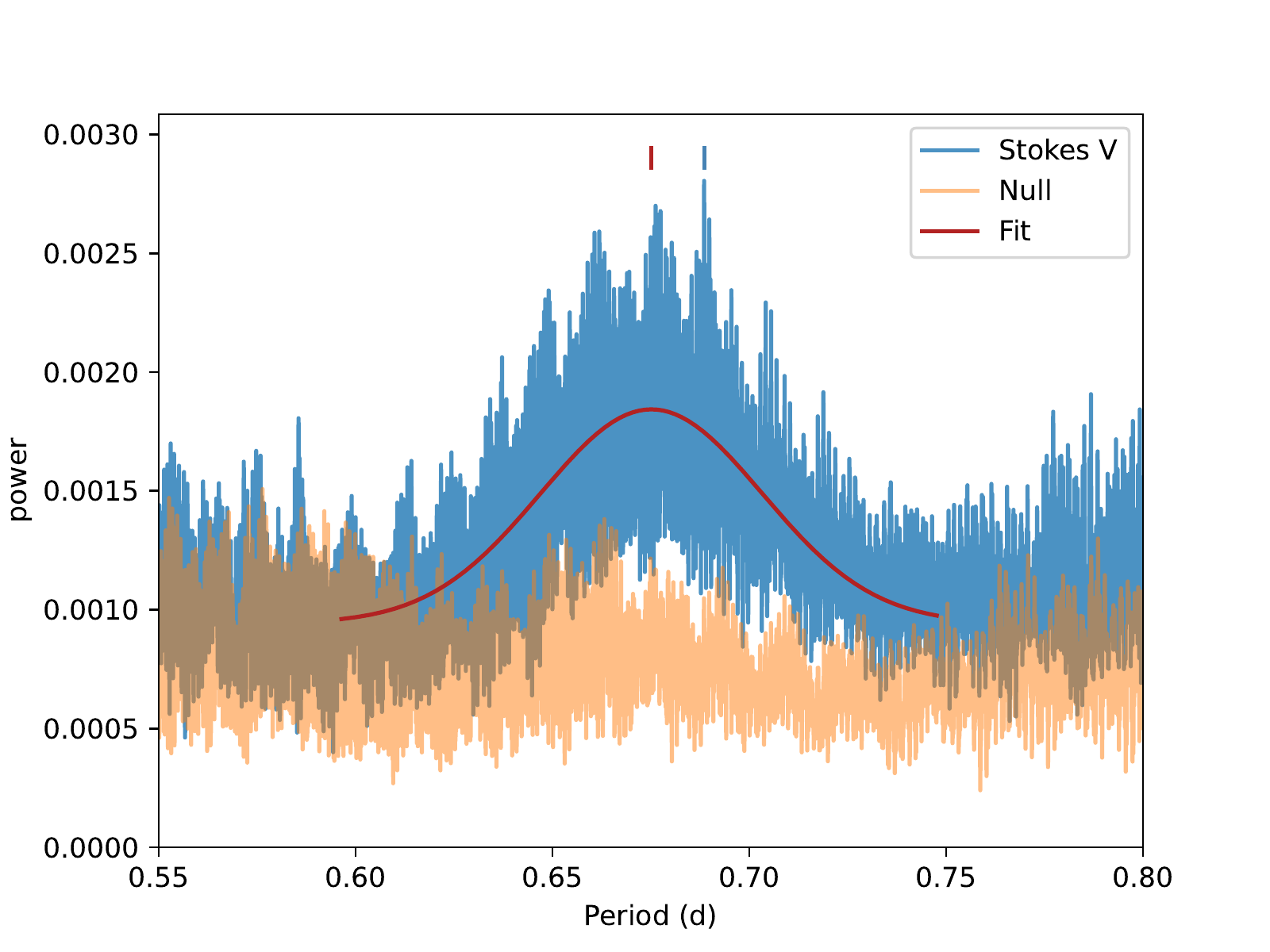}
\includegraphics[width=9cm]{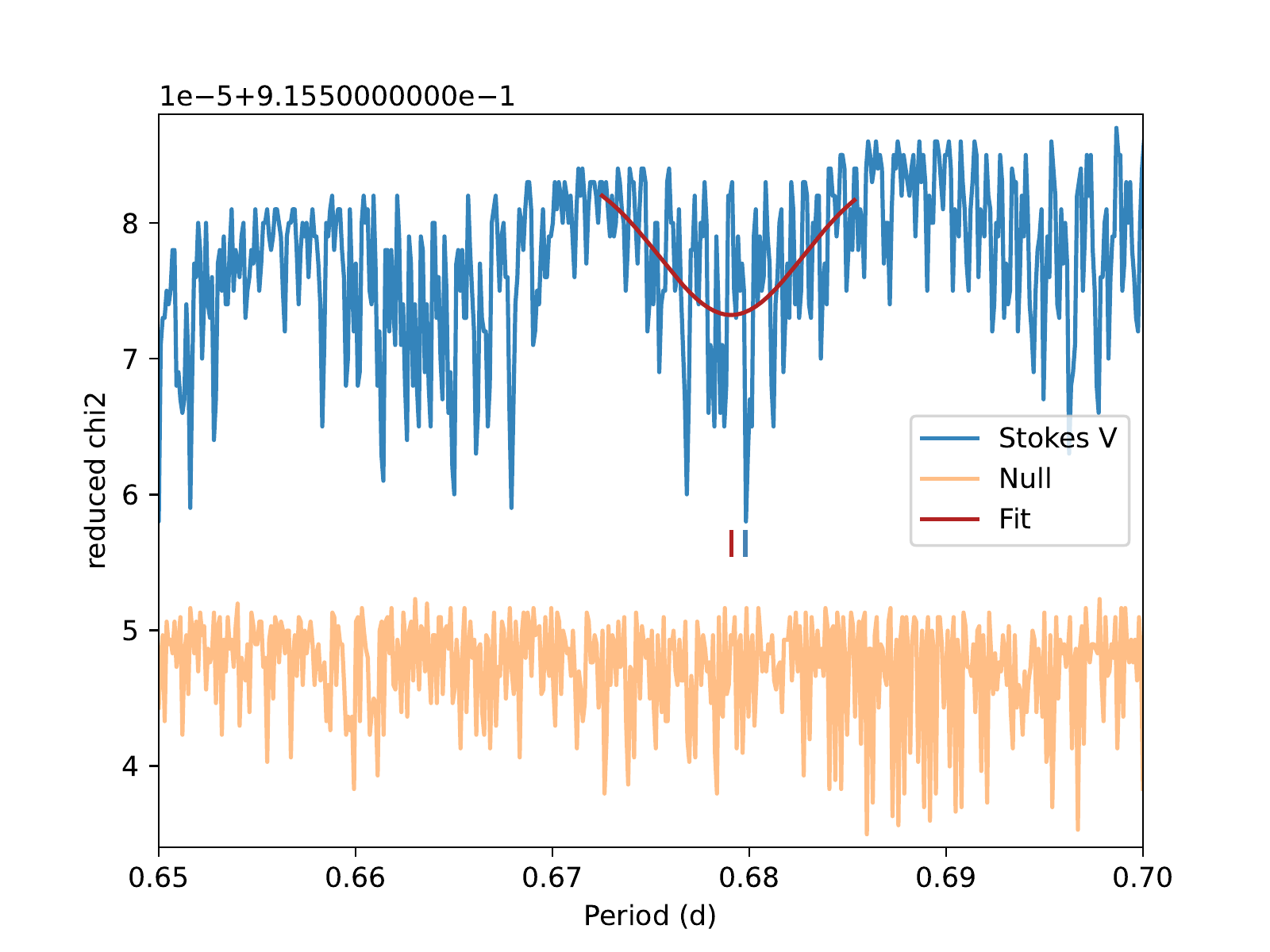} 
\caption{Period search using polarized data. Top: averaged periodograms obtained with Stokes V (blue) and with Null profiles (orange). A Gaussian fit is shown in red. Vertical ticks mark the best period found from the fit (red) and from the data (blue). Bottom: same, but using the reduced \kis\ of ZDI models.}
\label{fig:periodStokes}
\end{figure}

\begin{table}
\caption{Outcome of the period search. The first column is the selected subset. The second and third columns show, respectively, the period obtained in Stokes V profiles and from the ZDI model. The last column gives the reduced \kis\ of the ZDI inversion.} 
\centering
\begin{tabular}{c c c c}
\hline
 & \multicolumn{2}{c}{Period (d)} & \\
Year & V & ZDI ($\ell = 1$) & \kisr{}\ (ZDI) \\
\hline
2008 & 0.642 & 0.690 & 0.897 \\
2009 & 0.688 & 0.676 & 0.945\\
2010 & 0.691 & 0.664 & 0.905\\
2014 & 0.610 & 0.654 & 0.835\\
2018 & 0.667 & -- & 0.914\\
\hline
2008-2018 & 0.6751 (mean) & 0.6791 (mean) & 0.910\\ 
          & 0.688544  (best) & 0.6798 (best) & 0.910\\ 
\hline
\end{tabular}
\label{tab:periods}
\end{table}

The detailed characterization of the magnetic field geometry relies on a precise knowledge of the stellar rotation period. A value of 0.68~d was previously extracted by \cite{2015A&A...577A..64B} from a time series of high-resolution spectra. We perform here an independent period search, based on our own spectropolarimetric material and using two different approaches. We first make direct use of the Stokes V LSD profiles to look for periodic signatures in individual velocity bins. As a second option, we use the stellar rotation period as a free input parameter in the ZDI inversion to look for the value that optimizes the magnetic model.

\subsection{Periodicity in Stokes V profiles}

We constitute several time-series using individual velocity bins of the Stokes V LSD profiles, for each observing epoch. For each velocity bin, we extract a periodogram using the Astropy implementation of the Lomb Scargle algorithm \citep{1976Ap&SS..39..447L,1982ApJ...263..835S,2013A&A...558A..33A}. By doing so, we reproduce a procedure already employed by \cite{2018A&A...615A.116M}. 

The resulting periodograms are noise dominated, at least as long as we focus on individual velocity bins (not shown here). The situation can be improved by computing an averaged periodogram, obtained using all velocity bins within the line profile. The outcome for the 2008-2018 data set is plotted in the upper panel of Fig. \ref{fig:periodStokes}. Owing to the relatively large gaps between the observing epochs, a dense forest of peaks is observed. The highest peak is obtained at a period of 0.688544~d (Tab. \ref{tab:periods}). It is clear, however, that a broader peak is also visible. We estimated its location by a Gaussian fit (in red), obtaining a value of 0.6751~d. Both values are in overall agreement with \cite{2015A&A...577A..64B}. Running the same procedure with the Null profiles (orange curve in the same panel), we do not see any clear power peak, as expected with a Null profile containing noise.

Using the same procedure for each observing epoch, we obtain a series of values reported in Tab. \ref{tab:periods}. The subsets are dense enough to feature a single peak near the expected rotation period, although the time span of the subsets is most of the time of the order of a few weeks (up to 133 days at best, in 2009), sampling no more than a few hundreds of rotation phases. The degraded sampling, relative to the full data set comprising more than 2,000 rotation phases observed over a decade, leads to a dispersion of 0.03~d among this series of values, around an average of 0.660~d. Observations of 2010, which can be considered as the best of these sub-sets in terms of the number of observations and mean \sn, provide us with a period of 0.691~d. 

\subsection{ZDI period search}

The stellar rotation period is an input parameter of the ZDI code, as the tomographic modeling relies on the rotational phase distribution of the observations. By varying the input period, we obtain a series of magnetic models with different \kisr\ values, so that it is possible to identify the period optimizing the inversion process, which will provide us with the smallest \kisr. Period searches using ZDI were mainly performed on cool stars (e.g. \citealt{petit08,2017A&A...599A..72T,2021A&A...646A.130A}), and proved to be a powerful approach to isolating the stellar rotation signal. 

The same method was applied to the 2008 and 2009 epochs of Vega observations by \cite{petit10}. Their period value was, however, longer than the one reported by \cite{2015A&A...577A..64B}, and also longer than the period directly extracted from Stokes V profiles at the same epochs (Tab. \ref{tab:periods}). We repeated the measurements performed by \cite{petit10}, using our slightly adjusted ZDI model, and found a value compatible with their published one (around 0.72~d). Other epochs gave period values with a relatively large scatter, and sometimes no clear period at all (e.g. when all data were grouped together). A likely reason for this disappointing outcome is the large relative noise, making it nearly impossible to fully avoid to over-fit the data. Some input values of the rotation period may lead the noise pattern to mimic more efficiently a rotationally-modulated signal, leading to possible biases in our period estimate. The situation worsens when spherical harmonics modes with higher $\ell$ degrees are allowed in the ZDI model since the insertion of smaller-scale surface structures helps to misinterpret a white noise as an actual signal. An obvious workaround consists in restricting $\ell$ to values as small as possible. 

In Tab. \ref{tab:periods}, we show the periods obtained when we restrict the inversion to $\ell = 1$ (a justification of this strong additional constraint is given in Sec. \ref{sec:maps}). In this case, the period obtained from the 2008 and 2009 subsets ($\sim 0.69$ and $\sim 0.68$~d) become much closer to the ones directly extracted from the Stokes V profiles. \kisr\ variations obtained using the 2008-2018 data set as a whole are plotted in the lower panel of Fig \ref{fig:periodStokes}, showing again a dense distribution of \kisr\ minima, with a mean value of 0.6791~d (centroid of the red gaussian fit) and a best value of 0.6798~d. The same procedure, applied to the Null profiles, results in flatter \kisr\ fluctuations. We note that using all available data, the \kisr\ is progressively decreasing when smaller periods are probed (the onset of this decrease is visible on the left side of the plot), to the point that unrealistically short period values are actually favored. The likely reason for this outcome is that, even if the allowed magnetic geometry is forced to be very simple, shorter periods allow for misinterpretation of a fraction of the white noise as a rotational signal. One obvious conclusion is that the ZDI period search can become less stable when Zeeman signatures are vastly dominated by noise.

Using individual epochs, the values reported in Tab. \ref{tab:periods} have an average of 0.671~d, with a dispersion of 0.016~d, in overall agreement with the period search in Stokes V profiles. We could not identify a local \kisr\ minimum out of the 2018 data set. 

\section{Large-scale magnetic field geometry}
\label{sec:maps}

\begin{figure*} 
\centering
\includegraphics[width=19cm]{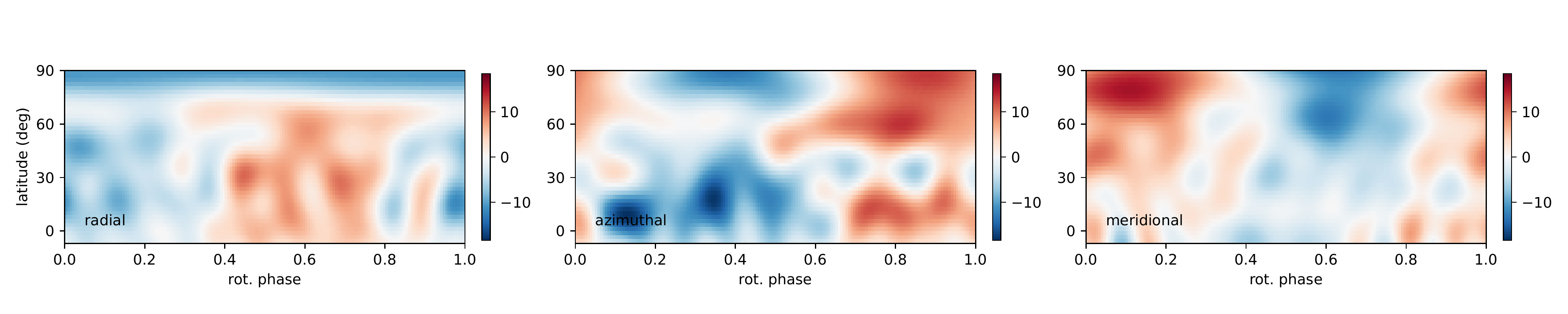} \\
\includegraphics[width=19cm]{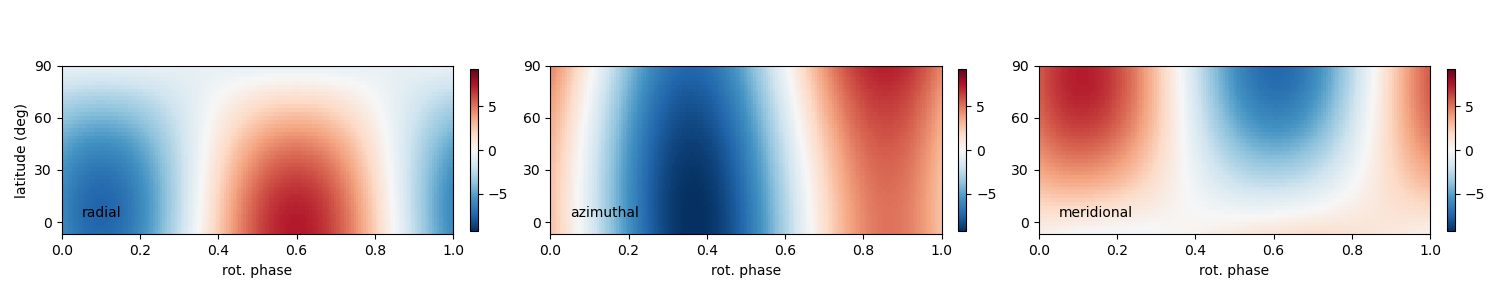} \\
\caption{Magnetic map obtained after grouping all data together (top). The three charts illustrate the radial, azimuthal and meridional components of the local magnetic vector. The field strength is expressed in gauss. The bottom map is obtained from the full data set, but limiting the spherical harmonics expansion to $\ell_{\rm max} = 1$. Note that the latitudinal extent of the maps is limited to the visible fraction of the stellar surface (at an inclination angle of 6.4\degr).}
\label{fig:maps}
\end{figure*}

The magnetic map obtained from the whole data set is shown in Fig. \ref{fig:maps}, while the series of maps derived from different subsets are displayed in Fig. \ref{fig:allmaps1}. As expected with a tomographic inversion of a polarized signal dominated by the noise, the finer details of the surface field (position and area of the smaller individual spots, local field strength) are not consistent between maps. We can, however, identify global similarities in this series of independent measurements.

In the equatorial projection adopted here, the most visible feature of the radial field component is a predominantly positive field between phases 0.4 and 0.8, at intermediate to low latitudes. Over the same latitude range, a negative polarity tends to dominate between phases 0.9 and 0.3, in phase opposition with the positive patch. We also report that the polar region is always covered by a radial field of negative polarity, with a strength of about -5~G. This polar spot is responsible for the phase-averaged signatures seen in Fig. \ref{fig:stokes}. Several features are also repeatedly observed in the azimuthal field component, with a negative polarity between phases 0.1 and 0.6 (at low to intermediate latitudes), replaced by a positive polarity over the rest of the rotation cycle (and closer to the pole). The meridional component is systematically stronger at higher latitudes and is characterized by the alternation between a positive polarity between phases 0.8 and 0.4, and a negative one for the rest of the phases. 

\begin{figure} 
\centering
\includegraphics[width=9cm]{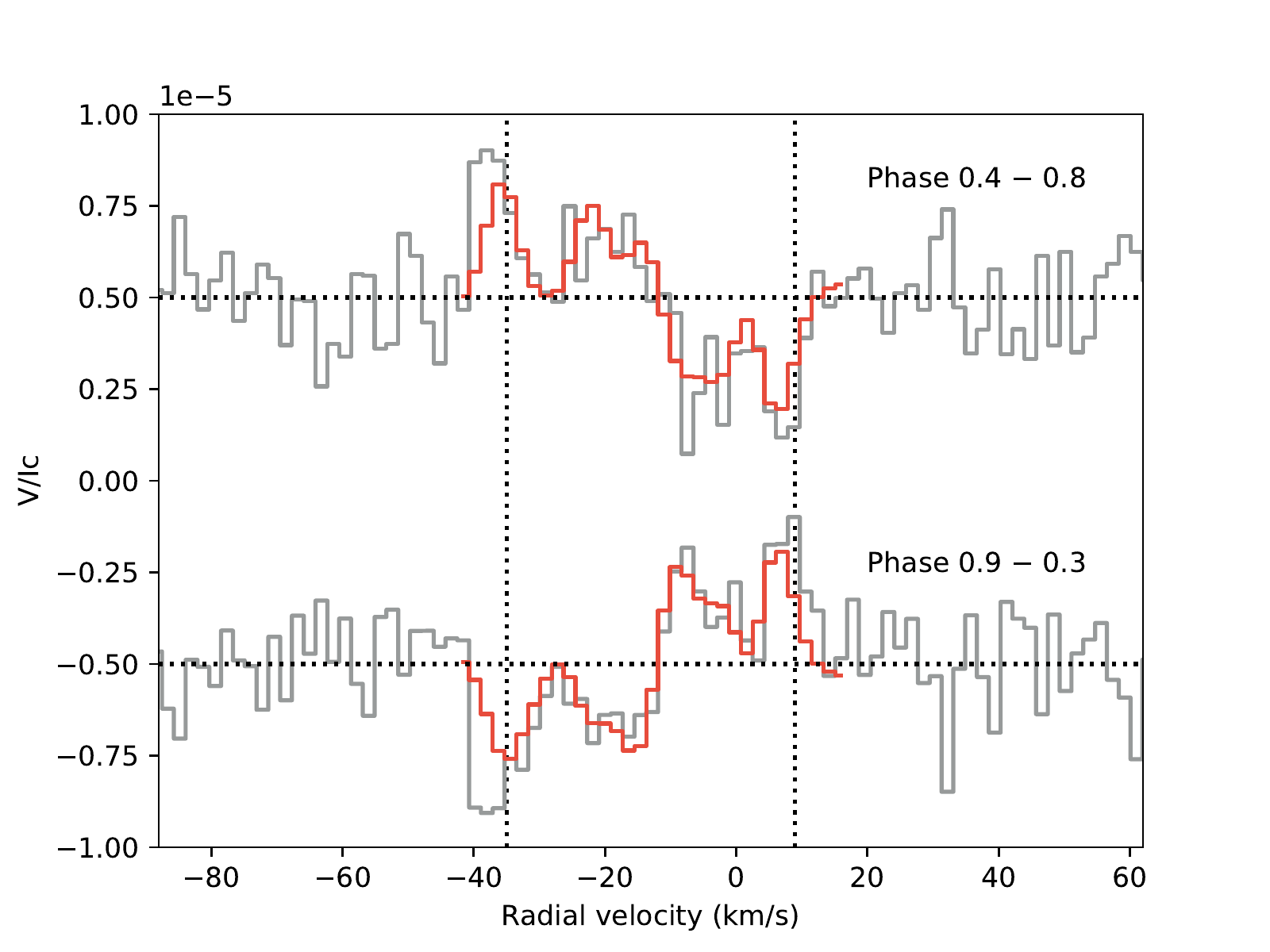}
\caption{Top plot: averaged LSD profiles obtained by selecting phases close to the positive pole of the dipole (phases between 0.4 and 0.8). The grey line shows observations, while the red line illustrates synthetic profiles produced by the ZDI model. The average of all available profiles (red curve in Fig. \ref{fig:stokes}) has been subtracted from observations and models to remove as much as possible the signature of the polar spot. Bottom plot: same, around the negative pole of the dipole (phases from 0.9 to 0.3). The dashed vertical lines mark the $\pm$~\vsin{}\ limit. The top and bottom plots were shifted vertically for a clearer display, with dashed horizontal lines showing the zero level.}
\label{fig:poles}
\end{figure}

Apart from the localized, radial polar spot, these field characteristics are reminiscent of the field distribution produced by a simple, very oblique dipole. To confirm this hypothesis, we ran another ZDI inversion of the whole data set, but this time we restrict the spherical harmonics to $\ell = 1$ (lowest panel of Fig. \ref{fig:maps}). By doing so, we reduce the spatial resolution of the map in this new inversion. A fair part of the characteristics detailed above survives in this simplified model, confirming that a weak dipole with a polar strength of $\sim 9$~G hosts a significant fraction of the surface magnetic energy. A toroidal component is also included in this inversion and accounts for a negligible 5\% of the overall surface magnetic energy. By selecting observations at the rotation phases of the $\ell = 1$ poles (Fig. \ref{fig:poles}), we obtain polarized line signatures with a polarity switch, as expected with an inclined dipole. However, we stress that such simple phase selection cannot isolate the specific Stokes V contribution of the dipole, because other magnetic components remain visible in the profiles. To filter out, as much as possible, the dominant signature of the polar spot, we have subtracted the average of all available profiles (which removes most of the polarized signature of the axisymmetric field component). The related amplitudes are about $0.6 \times 10^{-5}I_c$ peak-to-peak, which is about 1/3 of the signature amplitude produced by the polar spot. The dipole signatures are spread over the whole $\pm$~\vsin{} radial velocity span, while the polarized signature of the polar spot is confined in the core of the line.  

We note, however, that the $\ell = 1$ model is not sufficiently flexible to capture a number of smaller-scale structures. This is the case for the negative patch of radial field near the pole, totally absent from the map. This is also the case for other small features that can be observed with some consistency from year to year. For instance, there is evidence that what can be seen at low spatial resolution as the positive pole of the dipole is in fact a more complex structure, with at least two local maxima separated by about 0.2 rotation cycles. The azimuthal field projection is also more complex than its dipolar component, with negative spots repeatedly reconstructed in the large positive region. These specific features, among others, suggest that the magnetic topology of Vega is also characterized by some level of complexity. We stress, however, that a number of small magnetic spots are also less consistently reconstructed. Although some genuine variability of the surface field cannot be formally excluded, it is also likely that a fair part of these apparent changes may have a noise origin.     

\section{Discussion and conclusions}

The magnetic dipole reconstructed through tomographic inversion was not previously reported for Vega. The two magnetic maps of \cite{petit10} do show a different preferred field polarity for the two halves of the rotation cycle, but the evidence was judged too slim to be highlighted at that time. The additional epochs confirm this important property of the large-scale magnetic field of Vega. A global dipole was also reported for the Am star Alhena A, while the field topology of other magnetic Am stars is not known \citep{petit11,blazere16a}. The dipole obliquity in Ap stars was reported to depend on the rotation period \citep{2000A&A...359..213L,2002A&A...394.1023B}, with more oblique dipoles being observed more often in stars rotating in less than $\sim 25$~d, although  \cite{2019MNRAS.483.3127S} reached a less affirmative conclusion from a volume-limited sample of Ap stars. With its very fast rotation, the equatorial dipole of Vega tends to fall in line with the initially reported period dependence.  

A magnetic dipole inclined at 90\degr\ with respect to the rotation 
axis is equivalent to a $m=1$, $\ell=1$ spherical harmonic in the frame 
aligned with the rotation axis. The fact that azimuthal variations are 
dominated by the azimuthal number $m=1$ is compatible with the scenario 
of \cite{auriere07} where the non-Ap/Bp weak fields result from a 
Tayler instability of the magnetic configurations (see also \citealt{lignieres14,2015A&A...580A.103G,2015A&A...575A.106J}). The $m=1$ mode is 
the most unstable one for the Tayler instability. The  $m=1$, 
$\ell=1$ component of the magnetic field of Vega could then be the signature of a past Tayler instability.

Although it is very difficult to tell which small-scale magnetic regions are actually real in our reconstruction (except for the polar spot that is consistently recovered), we obtain repeated hints of a complex organization of the surface magnetism, with field strengths that can locally exceed the dipolar strength. This situation is reminiscent of the magnetic modeling of Alhena A, where the assumption of a pure dipole was not flexible enough to lead to a convincing magnetic model \citep{2020MNRAS.492.5794B}. The surface brightness of Vega was also reported to be a complex arrangement of dark and bright spots \citep{2017MNRAS.472L..30P}. A companion study (B\"ohm et al., in prep) will address the comparison between brightness and magnetic surface maps. At this stage, we can at least outline that the brightness maps of \cite{2017MNRAS.472L..30P} do not highlight an obvious large-scale brightness counterpart to the magnetic dipole. 

The main features of the magnetic topology (oblique dipole, polar spot) are remarkably stable over 10 years. Such stability is expected \citep{auriere07,2020ApJ...900..113J}, and actually observed \citep{silvester14}, in the presence of sufficiently strong fields, but is an interesting observation at such weak magnetic field strength. Spectroscopic time-series suggest that the local surface brightness may be affected by changes in the spot configuration over timescales as short as a few days \citep{2017MNRAS.472L..30P}. Such possible short-term variability would be very difficult to track using the polarimetric signal, since reconstructing magnetic maps with data restricted to individual nights is jeopardized by the noise, leaving us with large time gaps between successive maps. 

Other hints of short-term surface magnetic variability come from Alhena A, where the dipole was reported to be stable over several months, while smaller spots seemed to change over the same time scale \citep{2020MNRAS.492.5794B}. It was proposed by these authors that the smaller spots were dragged by surface flows and that a fair fraction of this drift could be modeled by a simple solar-like differential rotation law. We performed a similar search for a global surface shear using data taken in 2010 (since this sub-set provided us with an optimal phase sampling and \sn) but obtained inconclusive results. The absence of any noticeable differential rotation is consistent with negative results previously obtained for Vega from the cross-correlation of magnetic maps \citep{petit10} and brightness maps \citep{2017MNRAS.472L..30P}.  

\begin{acknowledgements}

This research made use of NASA's Astrophysics Data System Abstract Service.

\end{acknowledgements}

\bibliographystyle{aa}
\bibliography{vega.bib}

\begin{appendix}

\section{Histogram of phase coverage}

\begin{figure} 
\centering
\includegraphics[width=9cm]{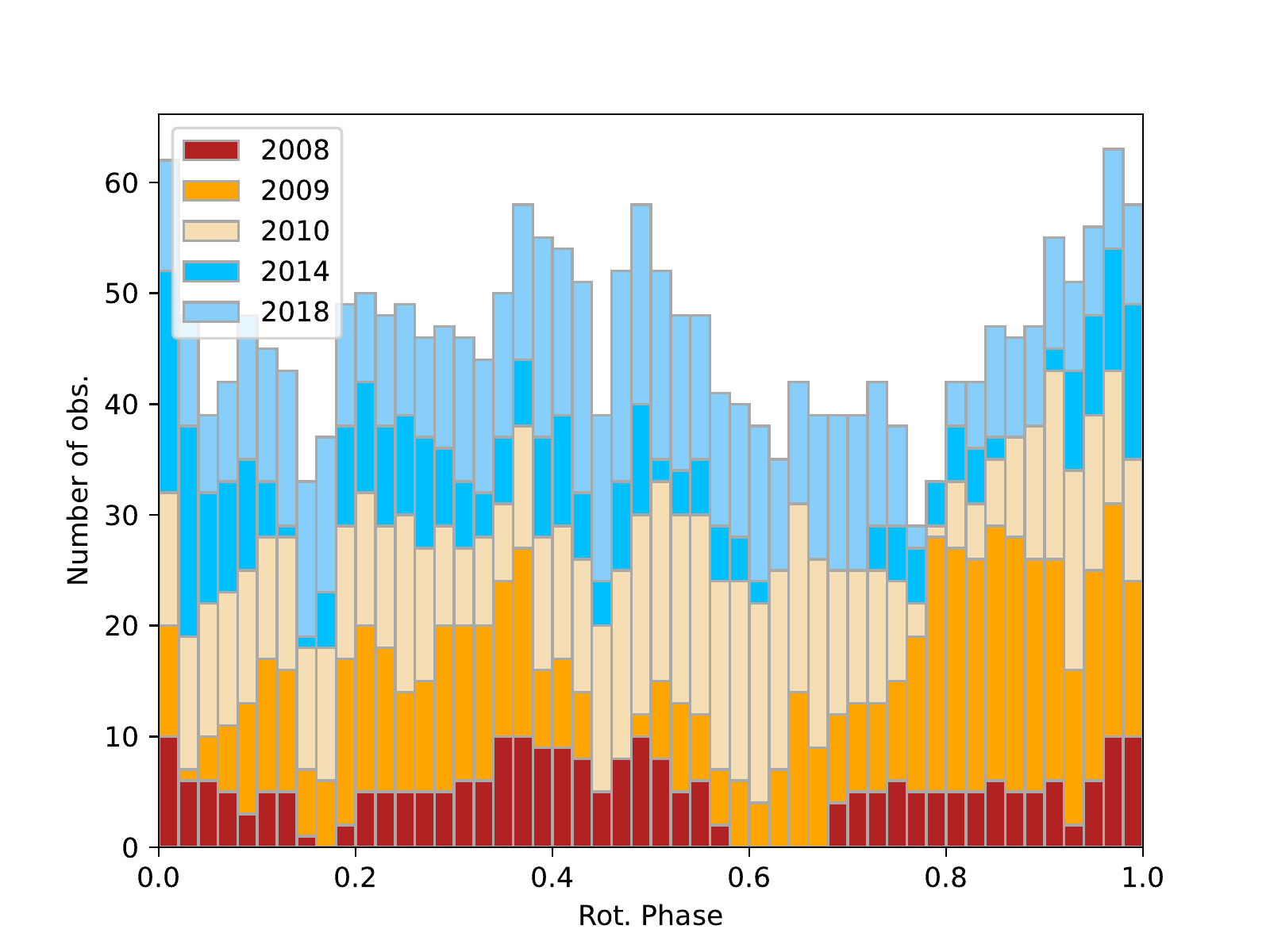}
\caption{Cumulative histogram of the available rotation phases, split by observing year. The ephemeris used to compute the phase is the same as the one presented in Sec. \ref{sec:maps}.}
\label{fig:phases}
\end{figure}

\section{Null LSD profiles}

\begin{figure} 
\includegraphics[width=9cm]{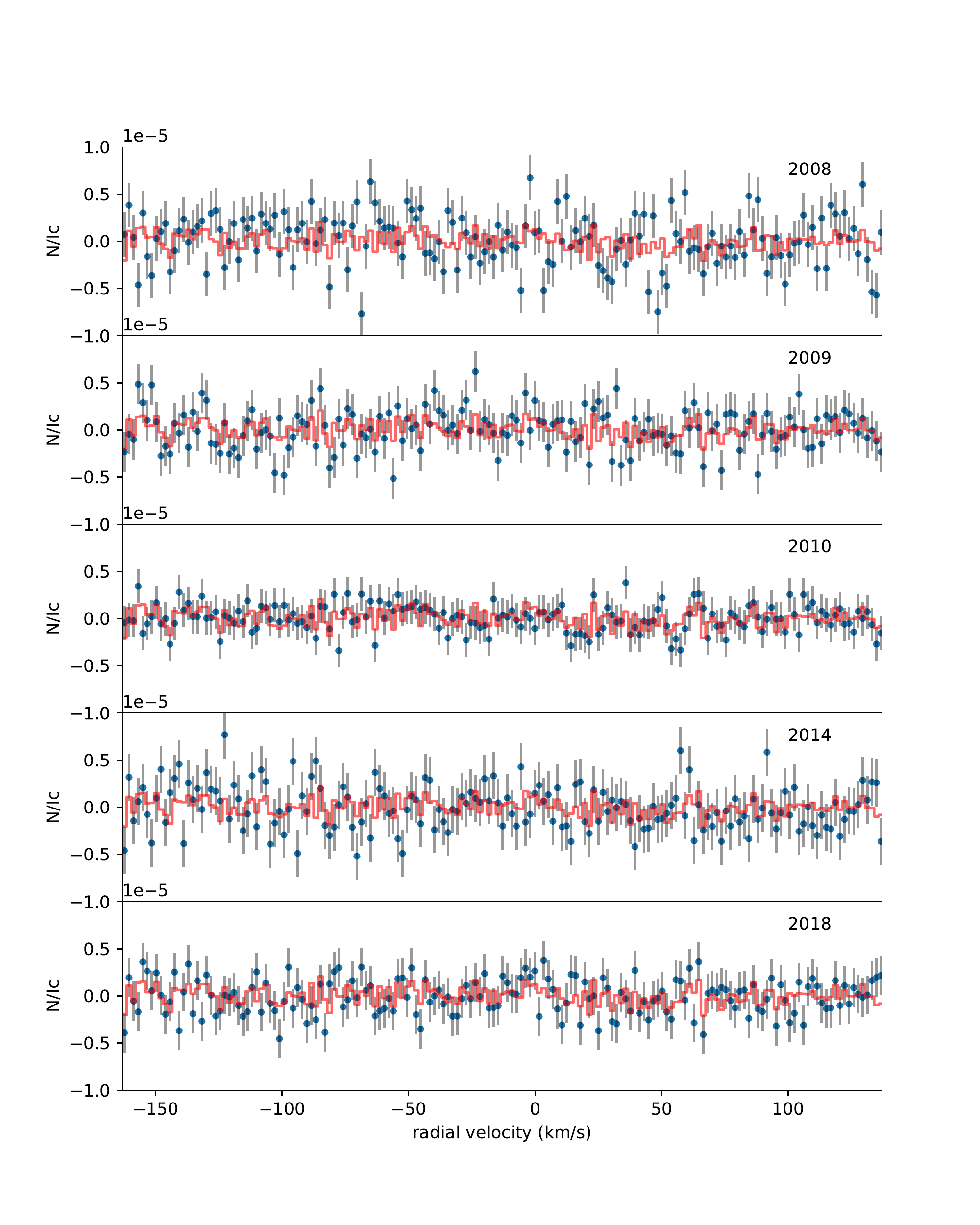} 
\caption{Same as Fig. \ref{fig:stokes}, but for the Null LSD profiles.}
\label{fig:null}
\end{figure}

\section{Magnetic maps}

\begin{figure*} 
\centering
\includegraphics[width=18cm]{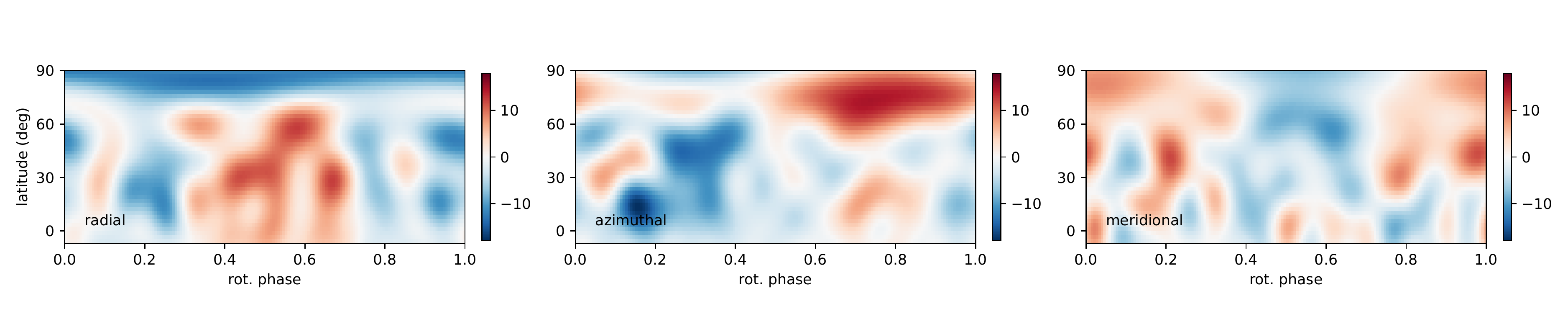} \\
\includegraphics[width=18cm]{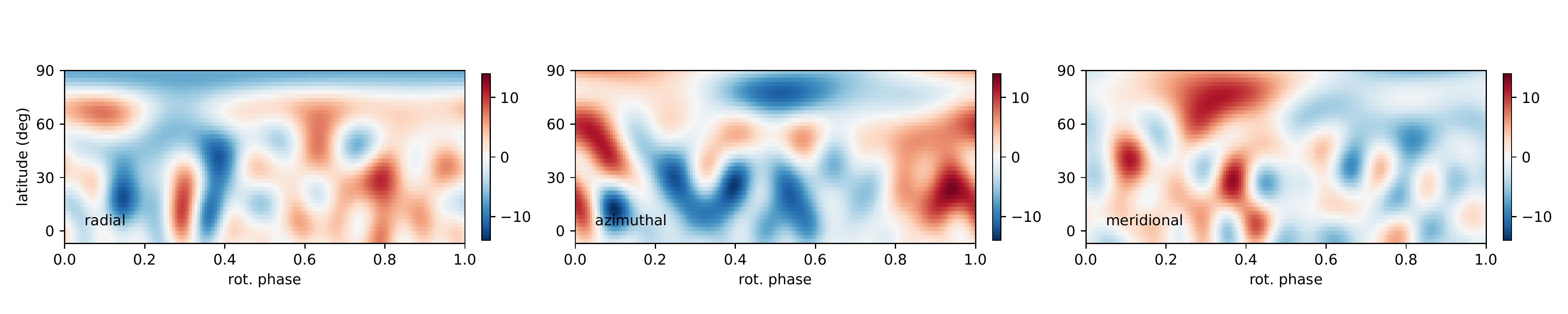} \\ 
\includegraphics[width=18cm]{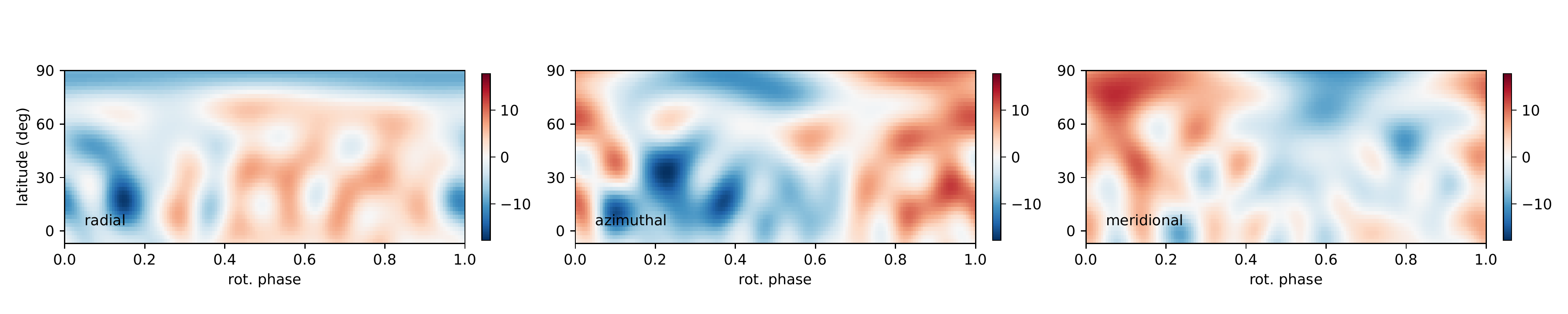} \\ 
\includegraphics[width=18cm]{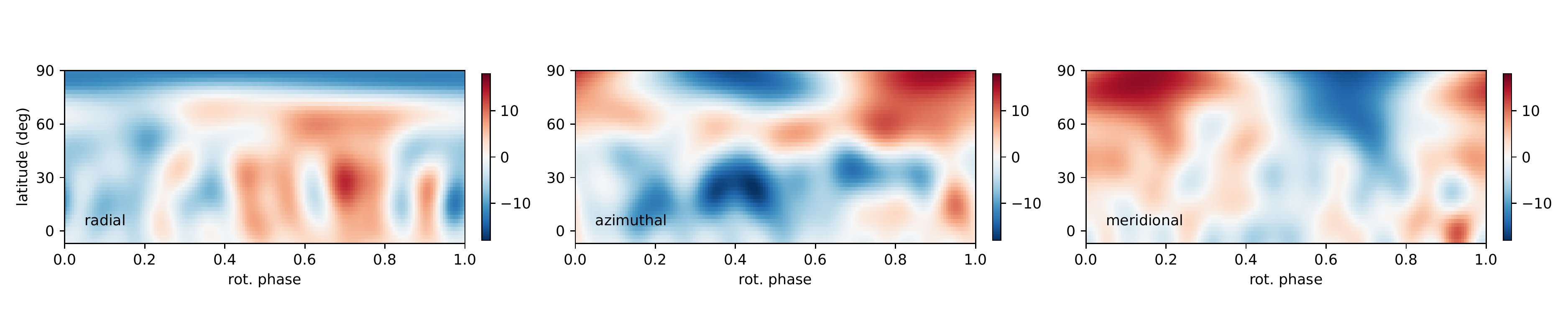} \\
\includegraphics[width=18cm]{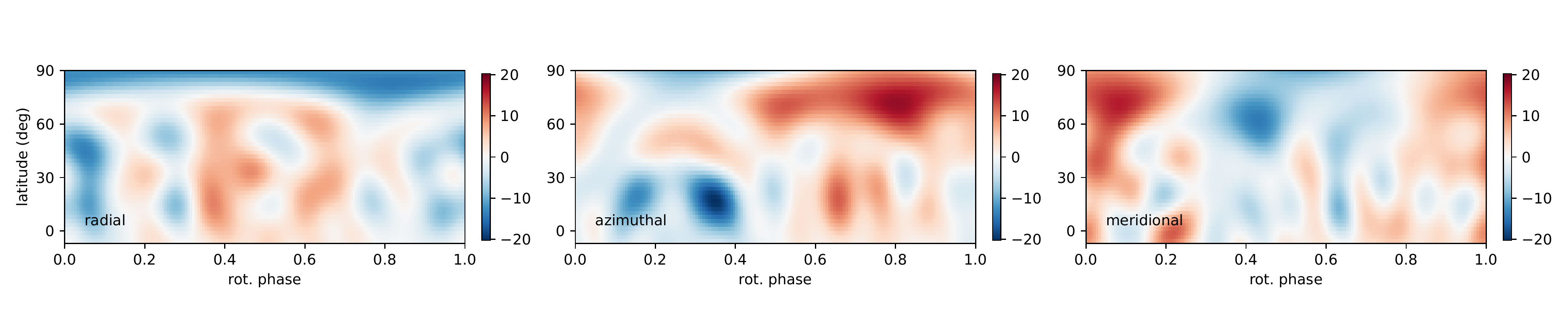} \\
\includegraphics[width=18cm]{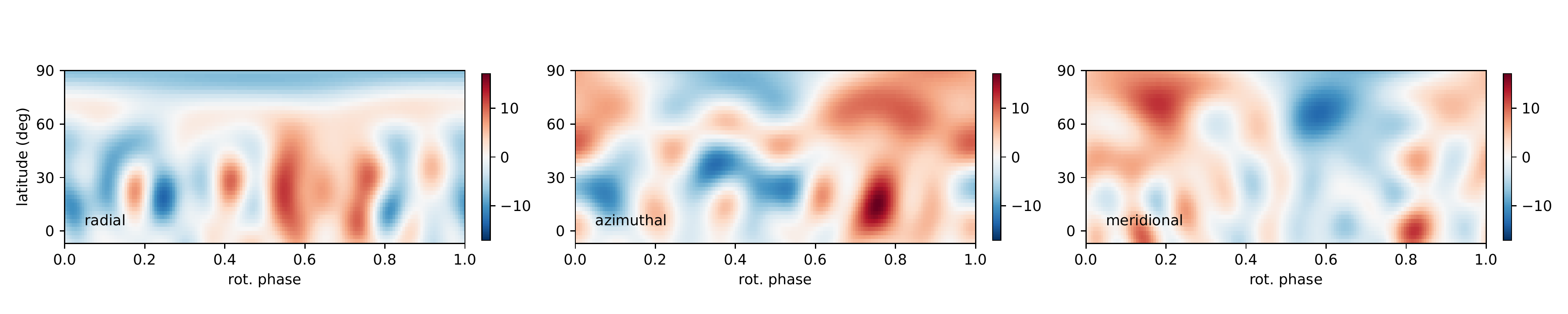} \\
\caption{Magnetic maps obtained for different subsets. Each line shows the radial (left), azimuthal (middle), and meridional (right) field components. Each horizontal strip displays the magnetic geometry reconstructed for one epoch, with from top to bottom data for 2008, 2009 (ESPaDOnS only), 2009 (NARVAL and ESPaDOnS), 2010, 2014, 2018. The field strength is expressed in gauss, and the color scales are identical for all maps, except for the bottom one.}
\label{fig:allmaps1}
\end{figure*}

\end{appendix}

\end{document}